\documentclass[a4paper,11pt]{article}
\usepackage{pos}
\usepackage{relsize}
\def\alphad{\alpha_{\raisebox{-1pt}{\tiny  D}}}

\title{Heaven and Earth: Connecting Jefferson Lab to the Cosmos}

\author*[a]{J. Piekarewicz}

\affiliation[a]{Department of Physics,
                    Florida State University,
	            Tallahassee, FL 32306-4350, 
	            USA}

\emailAdd{jpiekarewicz@fsu.edu}

\abstract{The nuclear equation of state (EOS) serves as the fundamental bridge between atomic nuclei and 
neutron stars---objects that differ in size by almost 20 orders of magnitude. Central to this connection is the 
nuclear symmetry energy, which controls both the neutron skin thickness of heavy nuclei and the radii of 
neutron stars. Recent electroweak experiments at Jefferson Lab, specifically PREX, have provided the 
cleanest terrestrial constraints on the EOS near saturation density. Complementary to neutron skins is the 
isovector giant dipole resonance, particularly the electric dipole polarizability. This contribution discusses 
the implications of these measurements on the nuclear EOS and the role of upcoming electroweak 
capabilities at Jefferson Lab in addressing their impact on the structure and composition of neutron stars.}

\FullConference{International Workshop on Low Energy Electron Positron Physics at Jefferson Lab (LEEPP2026)\\
23--27 March 2026\\ Thomas Jefferson National Accelerator Facility, Virginia, USA\\}

\begin{document}
\maketitle

\section{Introduction}

The quest to understand the structure of matter connects phenomena that span nearly twenty 
orders of magnitude in length scale. At one extreme lie atomic nuclei, femtometer-sized many-body 
quantum systems that can be probed with exquisite precision using low-energy electron scattering. 
At the other extreme lie neutron stars, kilometer-sized remnants of stellar evolution that contain matter 
compressed to densities far exceeding those found inside atomic nuclei. Remarkably, both systems 
are governed by the same underlying nuclear interactions. This intimate connection between terrestrial
 laboratory experiments and astrophysical observations has transformed nuclear physics into a truly 
 multidisciplinary field in the emerging era of multi-messenger astronomy. In this contribution, I discuss 
 how precision measurements at Jefferson Lab, particularly parity-violating electron scattering experiments 
 that probe the neutron distribution of nuclei, provide critical information on the nuclear equation of state 
 and its symmetry energy, thereby linking laboratory measurements here on Earth to some of the most 
 extreme objects in the universe. I further argue that future high-precision measurements of the nuclear 
 dipole response can complement parity-violating electron scattering by extending sensitivity to exotic 
 nuclei with large neutron-proton asymmetries.

Among the central questions of modern nuclear astrophysics are the origin of the heavy elements 
and the nature of matter under the most extreme conditions found in the universe. Framed within 
the context of the Long Range Plans for Nuclear Science\,\cite{LRP2015,LRP2023}, these grand 
challenges may be posed as follows: How were the heavy elements, from iron to uranium, forged 
in the cosmos? And what novel forms of matter emerge under conditions of extreme density and 
temperature? Although these questions concern astrophysical phenomena on the largest scales, 
their answers rely on the microscopic properties of strongly interacting matter and the nuclear forces 
that govern its behavior.

In the particular case of neutron stars, the theoretical framework that connects microscopic nuclear
 physics to macroscopic observables is provided by the nuclear equation of state (EOS). The EOS 
 determines the pressure generated by matter at a given density and composition and therefore 
 controls how matter responds to compression. Through the Tolman-Oppenheimer-Volkoff equations, 
 the EOS serves as the primary physical input governing the structure and global properties of neutron 
 stars\,\cite{Tol39_PR55,Opp39_PR55}. It is through this remarkable connection between the properties
 of finite nuclei measured in terrestrial laboratories and the bulk properties of neutron stars inferred
 from astronomical observations that the theme of  ``Heaven and Earth'' emerges most naturally.

\section{The Nuclear Equation of State}
\label{Sec:EOS}

As mentioned in the Introduction, the nuclear equation of state (EOS) provides the microscopic foundation 
for understanding the structure and composition of neutron stars. Because the Fermi temperature of dense 
matter is far larger than the actual temperature of the star, neutron-star matter may be regarded as effectively 
``cold'', so the EOS depends only on the neutron ($n_{n}$) and proton ($n_{p}$) densities. Equivalently---and 
more commonly used in the literature---the energy per baryon may be expressed in terms of the conserved 
baryon density $n$ and the neutron-proton asymmetry $\alpha$ defined as follows:
\begin{subequations}
\begin{align}
  n & = n_{n} + n_{p}, \\
 \alpha &=\frac{n_{n}-n_{p}}{n_{n}+n_{p}}.
\end{align}
\end{subequations}
That is,
\begin{equation}
  \frac{E}{A}(n,\alpha) -\!M \equiv {\cal E}(n,\alpha) = {\cal E}_{\rm SNM}(n)
                  + \alpha^{2}{\cal S}(n) + {\cal O}(\alpha^{4}) \,,
 \label{EOS}
\end {equation}
where ${\cal E}_{\rm SNM}(n)\!=\!{\cal E}(n,\alpha\!\equiv\!0)$ is the energy per nucleon of symmetric 
nuclear matter (i.e., $n_{n}\!=\!n_{p}$) and the symmetry energy ${\cal S}(n)$ represents the leading 
correction associated with the neutron-proton asymmetry. To an excellent approximation, the symmetry 
energy may be regarded as the energy cost required to convert symmetric nuclear matter into pure 
neutron matter:
\begin{equation}
 {\cal S}(n) \equiv \frac{1}{2}\left(\frac{\partial^{2}{\cal E}(n,\alpha)}{\partial\alpha^{2}}\right)_{\!\alpha=0}
 \!\approx\!\Big[{\cal E}(\rho,\alpha\!=\!1) \!-\! {\cal E}(\rho,\alpha\!=\!0)\Big] \;.
 \label{SymmE}
\end {equation}

It is customary to characterize the behavior of both symmetric nuclear matter and the symmetry energy 
in terms of a few bulk parameters defined at saturation density 
$n_{0}\!\approx\!0.15\,{\rm fm}^{-3}$\,\cite{Piekarewicz:2008nh}:
\begin{subequations}
\begin{align}
 & {\cal E}_{\rm SNM}(n) = \mathlarger{\varepsilon}_{0} + \frac{1}{2}Kx^{2}+\ldots ,\label{EandSa}\\
 & {\cal S}(\rho) = J + Lx + \frac{1}{2}K_{\rm sym}x^{2}+\ldots ,\label{EandSb}
\end{align} 
\end{subequations}

where $x\!=\!(n-n_{0})/3n_{0}$ is a dimensionless parameter that quantifies deviations from saturation density. Here 
$\mathlarger{\varepsilon}_{0}$ and $K$ denote the binding energy per nucleon and incompressibility coefficient of 
symmetric nuclear matter, while $J$, $L$, and $K_{\rm sym}$ characterize the value, slope, and curvature of the 
symmetry energy at saturation. Unlike symmetric nuclear matter which saturates at $n_0$, the symmetry energy 
has a nonvanishing slope at saturation. This slope determines the pressure of neutron-rich matter near saturation 
density and thereby influences both the neutron skin of heavy nuclei and the structure of neutron stars. Indeed, 
assuming the validity of the approximation given in Eq.(\ref{SymmE}), $L$ is directly proportional to the pressure of 
pure neutron matter at saturation density:
\begin{equation}
   P_{\rm PNM}(n_{0}) \approx \frac{1}{3}n_{0}L \;.
 \label{PvsL}
\end{equation}
This simple relation captures the essence of the connection between ``Heaven and Earth'': the same pressure that 
pushes neutrons toward the surface of a heavy nucleus also supports a neutron star against gravitational collapse.

Although these bulk parameters provide an intuitive characterization of the EOS, they are not themselves physical 
observables that can be measured directly in the laboratory. Instead, they must be inferred from observables that 
are sensitive to the underlying distance structure of the nuclear interaction. Among these, the neutron-skin 
thickness of neutron-rich nuclei has emerged as a particularly powerful probe of the symmetry energy. As discussed 
in the next section, precision measurements of the neutron skin of ${}^{208}$Pb at Jefferson Lab provide important 
constraints on the slope parameter $L$ and, consequently, on the pressure of neutron-rich matter near saturation 
density.

However, unlike infinite asymmetric nuclear matter, where the baryon density $n$ and the neutron-proton asymmetry 
$\alpha$ may be regarded as independent thermodynamic variables, the composition of cold, catalyzed neutron-star 
matter is determined by the conditions of beta equilibrium and charge neutrality. At each baryon density, the equilibrium 
composition is obtained by minimizing the total energy of a system consisting of neutrons, protons, and a neutralizing 
Fermi gas of leptons (electrons and, above threshold, muons). This requires that the chemical potentials satisfy
\begin{equation}
  \mu_{n} = \mu_{p} + \mu_{e} \hspace{4pt}{\rm and}\hspace{4pt} \mu_{e} = \mu_{\mu}. 
 \label{ChemEq1}
\end{equation}
where the first condition expresses equilibrium under weak interactions and the second ensures equilibrium between 
electrons and muons.

To gain physical insight into the composition of neutron-star matter, consider the idealized limit in which muons are 
absent and the density is sufficiently high that the masses of all constituents may be neglected. In this case, beta 
equilibrium together with charge neutrality yields a proton fraction of $Y_{p}\!=\!1/9$, corresponding to a neutron-proton 
asymmetry of $\alpha\!=\!7/9$. Thus, it is primarily the requirement of charge neutrality---rather than the nuclear interaction 
itself---that is responsible for the pronounced neutron excess characteristic of neutron-star matter.

\section{Electroweak Probes of Neutron Densities}

In this section we discuss how precision measurements of the neutron skin of ${}^{208}$Pb at Jefferson Lab provide 
powerful constraints on the equation of state of neutron-rich matter. As discussed in the previous section, the slope 
of the symmetry energy $L$ determines the pressure of neutron-rich matter near saturation density. Because this same 
pressure governs the spatial distribution of neutrons in heavy nuclei, measurements of neutron skins offer a unique 
terrestrial probe to the structure of neutron stars.

Parity-violating electron scattering (PVES) was proposed more than three decades ago as a clean and largely 
model-independent probe of neutron densities\,\cite{Donnelly:1989qs}. Since then, a sustained experimental 
effort at Jefferson Lab has established PVES as a powerful tool for studying neutron-rich 
matter\,\cite{Abrahamyan:2012gp,Adhikari:2021phr,Adhikari:2022kgg}. Interest in such measurements was 
further amplified by the realization that neutron densities, and in particular the neutron-skin thicknesses, provide 
valuable constraints on the equation of state of neutron-rich matter and the structure of neutron 
stars\,\cite{Horowitz:2000xj}.

The parity-violating asymmetry arises from the interference between electromagnetic scattering mediated by photon 
exchange and weak neutral-current scattering mediated by the $Z^0$ boson. In Born approximation, the asymmetry 
is given by
\begin{equation}
  A_{PV}(Q^{2}) = \frac{G_{\!F}Q^{2}}{4\pi\alpha\sqrt{2}}
  \frac{Q_{\rm wk}F_{\rm wk}(Q^{2})}{ZF_{\rm ch}(Q^{2})},
\label{APVb}
\end{equation}
where $Q^{2}$ is the four-momentum transfer, $\alpha$ the fine-structure constant, $G_{F}$ the Fermi constant, 
$Z$ the nuclear charge, and $Q_{\rm wk}$ the weak nuclear charge. The relevant nuclear-structure information 
is encoded in the charge and weak form factors, $F_{\rm ch}(Q^2)$ and $F_{\rm wk}(Q^2)$, respectively. Because
charge densities are already known with high precision from conventional electron-scattering experiments, a 
measurement of $A_{PV}$ provides direct access to the weak form factor and, therefore, to the neutron distribution. 
Although enormously challenging, such measurements offer a unique electroweak probe of neutron-rich matter that 
is largely free from the model uncertainties associated with hadronic reactions, thereby providing a clean and 
model-independent determination of neutron densities in atomic nuclei.

\begin{figure}[ht]
\centering
\bigskip
\includegraphics[width=\linewidth]{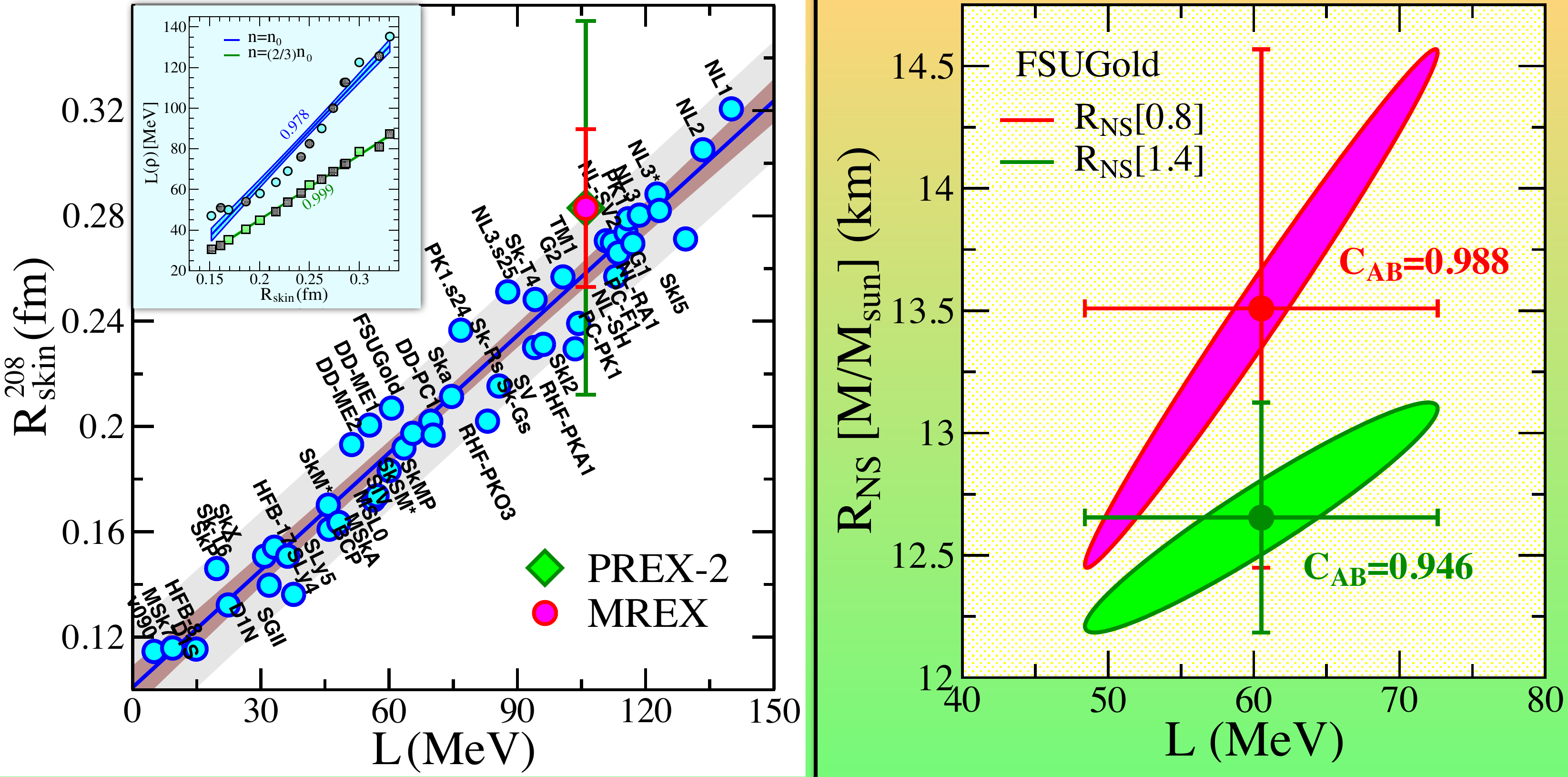}
\caption{Left: Correlation between the neutron-skin thickness of ${}^{208}$Pb and the slope of the symmetry 
energy $L$ predicted by a representative set of non-relativistic and relativistic energy density functionals. The 
PREX measurement and the projected MREX uncertainty are also displayed. The inset shows that the correlation 
becomes even stronger when the slope is evaluated at approximately two-thirds of nuclear saturation density. 
Right: Statistical correlation between $L$ and the radii of 0.8- and 1.4-solar-mass neutron stars predicted by the 
FSUGold energy density functional, illustrating the common dependence of nuclear and neutron-star observables 
on the pressure of neutron-rich matter in the vicinity of saturation density.}
\label{Fig1}
\end{figure}

Figure\,\ref{Fig1} illustrates the intimate connection between laboratory measurements of finite nuclei and the structure 
of neutron stars. The left-hand panel displays the correlation between the neutron-skin thickness of ${}^{208}$Pb and 
the slope of the symmetry energy at saturation density\,\cite{Brown:2000,Furnstahl:2001un,RocaMaza:2011pm} predicted 
by a representative set of both non-relativistic and relativistic energy density functionals. Despite the diversity of the 
underlying theoretical approaches, the predicted correlation is remarkably strong, demonstrating that the neutron skin 
provides a sensitive probe of the density dependence of the symmetry energy. The figure also includes the PREX 
measurement together with the projected precision anticipated from the upcoming MREX experiment. In turn, the 
inset illustrates that the correlation becomes even tighter when the symmetry-energy slope is evaluated at 
approximately two-thirds of nuclear saturation density. 

The right-hand panel demonstrates that the same parameter governing the neutron skin also strongly influences neutron-star 
radii. Shown are the statistical correlations between $L$ and the radii of 0.8- and 1.4-solar-mass neutron stars predicted by the 
FSUGold energy density functional. Although the neutron skin of ${}^{208}$Pb measures only a few tenths of a femtometer 
while neutron-star radii extend over roughly 13\,km---a difference approaching twenty orders of magnitude---both observables 
are controlled by the pressure of neutron-rich matter near nuclear saturation density. This remarkable correspondence provides 
one of the clearest examples of the deep connection between terrestrial nuclear experiments and astrophysical observations.

\section{Isovector Giant Dipole Resonance}
\label{IVGDR}

Although parity-violating electron scattering provides the cleanest determination of neutron densities, complementary 
information on the symmetry energy can be obtained from the nuclear dipole response. We therefore devote this section 
to the role of the isovector giant dipole resonance in constraining the equation of state of neutron-rich matter. In a simple 
macroscopic picture, the isovector giant dipole resonance (GDR) may be viewed as an out-of-phase collective oscillation 
of neutrons against protons. As these two quantum fluids are displaced in opposite directions from their equilibrium 
positions, the symmetry energy acts as the restoring force, making this fundamental mode highly sensitive to its density 
dependence. However, because the energy-weighted sum rule is largely model independent, this moment of the strength
distribution is insensitive to the symmetry energy\,\cite{Piekarewicz:2010fa}.

Instead, the inverse energy-weighted sum $m_{-1}$, has emerged as a powerful isovector 
indicator\,\cite{Reinhard:2010wz,Piekarewicz:2010fa,Piekarewicz:2012pp,Roca-Maza:2013mla}. Moreover, because the 
low-energy part of the dipole response dominates $m_{-1}$, this moment is particularly well suited for studying the so-called 
pygmy dipole resonance (PDR), which is itself closely connected to the development of a neutron-rich 
skin\,\cite{Piekarewicz:2006ip,Savran:2013bha}. The electric dipole polarizability $\alpha_D$ is simply related to $m_{-1}$ 
through
\begin{equation}
\alphad = \frac{8\pi}{9}e^{2}m_{-1}.
\label{DipPol}
\end{equation}

High-resolution measurements of the dipole response of ${}^{208}$Pb have been completed at RCNP in 
Japan\,\cite{Tamii:2011pv,Poltoratska:2012nf,Tamii:2013cna}. Although the response was measured using inelastic proton 
scattering, a major advantage of these experiments is that, at forward angles, Coulomb excitation dominates the reaction 
mechanism, allowing the extraction of $\alpha_D$ with minimal strong-interaction uncertainties. Complementing these 
stable-target measurements are studies of the electric dipole response of neutron-rich unstable nuclei at 
GSI\,\cite{Klimkiewicz:2007zz,Adrich:2005,Wieland:2009,Rossi:2012ew}. Such experiments---both at Jefferson Lab
and elsewhere---will continue to be enormously valuable because the emergence of low-energy dipole strength has 
been shown to correlate strongly with the development of a neutron-rich skin, thereby providing additional insight into 
the density dependence of the symmetry energy\,\cite{Piekarewicz:2006ip}.

\begin{figure}[ht]
\begin{center}
\includegraphics[width=0.65\linewidth]{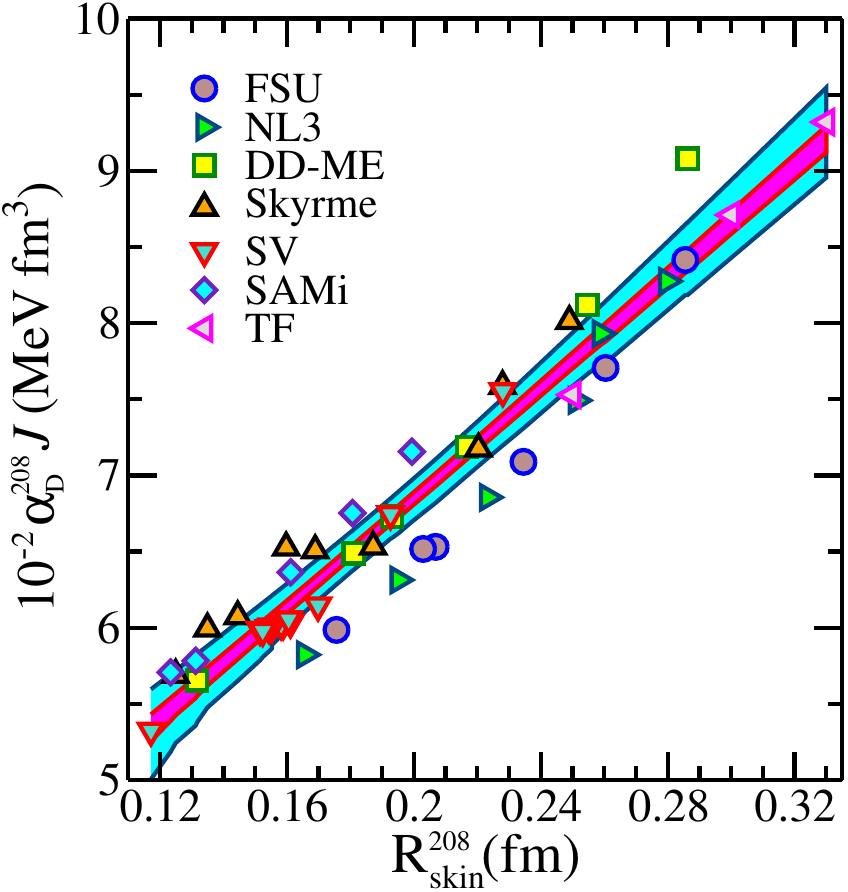}
\caption{Predictions from a representative set of relativistic and non-relativistic energy density functionals
for the product of the electric dipole polarizability of ${}^{208}$Pb and the symmetry energy at saturation 
density, $\alpha_{D}J$. These results were originally reported in Ref.\cite{Roca-Maza:2013mla}.}
\label{Fig2}
\end{center}
\end{figure}

Although the results displayed in Fig.\ref{Fig2} are obtained from fully self-consistent microscopic calculations, valuable 
physical insight emerges from the liquid-droplet model\,\cite{Satula:2005hy,Roca-Maza:2013mla}. In particular, the droplet 
model predicts that the product $\alpha_{D}J$, rather than the dipole polarizability alone, should exhibit the strongest 
correlation with the density dependence of the symmetry energy. Figure\,\ref{Fig2} confirms this expectation: predictions 
from a broad collection of relativistic and non-relativistic energy density functionals collapse onto an almost universal linear 
relation between the neutron-skin thickness of ${}^{208}$Pb and $\alpha_{D}J$. This remarkable robustness reflects the 
fact that both the neutron-skin thickness and the electric dipole polarizability probe the pressure generated by the symmetry 
energy at subsaturation densities, where the nuclear surface is formed.

\section{Conclusion}
In this proceedings we have explored the profound connection between ``Heaven and Earth'' through the nuclear 
equation of state and, in particular, through the density dependence of the nuclear symmetry energy. Whereas a 
precision measurement of the neutron-skin thickness provides a powerful constraint on the slope parameter $L$, 
its combination with an accurate determination of the electric dipole polarizability enables a simultaneous 
determination of both the magnitude of the symmetry energy at saturation, $J$, and its density dependence. In 
this way, the complementary experimental programs at Jefferson Lab, RCNP, and future rare-isotope facilities 
will continue to provide increasingly stringent constraints on the isovector sector of the nuclear equation of state, 
further strengthening the connection between laboratory measurements of finite nuclei and the structure and 
composition of neutron stars.

As Jefferson Lab enters a new era of high-precision electroweak measurements, the anticipated MREX experiment 
at Mainz and FRIB's expanding program on neutron-rich nuclei will provide powerful and complementary probes of 
the symmetry energy. Together with advances in nuclear theory and multi-messenger astrophysics, these efforts 
promise increasingly stringent constraints on the equation of state of dense matter. By connecting laboratory 
measurements of finite nuclei to the structure and composition of neutron stars, they will continue to strengthen 
the remarkable bridge between ``Heaven and Earth''.

\begin{acknowledgments}\vspace{-10pt}
This material is based upon work supported by the U.S. Department of Energy Office of Science, 
Office of Nuclear Physics under Award Numbers DE-FG02-92ER40750.
\end{acknowledgments}



\end{document}